\documentclass[doublecol]{epl2} 

\usepackage{amsmath}
\usepackage{amsfonts}
\usepackage{amssymb}
\usepackage{cite}
\usepackage{hyperref}

\makeatletter
\let\epl@pacsmissing\@empty
\let\epl@pacsset\@empty
\renewcommand{\pacs}[2]{}
\makeatother

\graphicspath{{figures/}{./}}

\title{Emergent Self-Organisation of Intelligent Active Particles}
\shorttitle{Self-Organization of Intelligent Active Particles} 

\author{Priyanka Iyer\inst{1} \and Segun Goh\inst{1,2} \and Gerhard Gompper\inst{1}}
\shortauthor{P.~Iyer \etal}

\institute{                    
  \inst{1} Theoretical Physics of Living Matter, Institute for Advanced Simulation, Forschungszentrum J\"ulich,
           D-52425 J\"ulich, Germany \\
  \inst{2} School of Liberal Studies, Sejong University, 05006 Seoul, Korea
}

\abstract{Intelligent active particles are characterized by self-propulsion, directional sensing of their environment,  information processing, decision making and goal-oriented self-steering. 
This implies, in particular, the prevalence of  non-reciprocal interactions, and the importance of information propagation through agent groups. 
Examples include biological systems (cells, insects, birds, fish, pedestrians) as well as engineered systems (nano- and microbots). As many agents move in an aqueous medium, hydrodynamic interactions strongly affect the dynamics. The emergent dynamics includes
the formation of swarms and flocks, predator-prey behavior, and the navigation in complex environments.
}

\begin{document}
\maketitle

\section{Introduction}
Motile active matter encompasses self-propelled particles and microswimmers, which due to their non-equilibrium nature display many novel types of dynamics and collective behaviors \cite{elgeti2015_review, bechinger2016active}. However, 
motile biological organisms can also sense their environment, process this information, and make decisions about favorable active responses. This applies to organisms over many length scales, from
large animals and humans down to the microscopic size of cells and bacteria. Similar principles are
also relevant for the development of microrobotic systems \cite{tsang2020_roads, gompper2025_roadmap_short, loewen2026_intelligentparticles, ju2025_Technology_short}. These systems are now often classified as
``intelligent active matter" \cite{kaspar2021_intelligent, baulin2025_embodied_short, loewen2026_intelligentparticles}.

Since humans and humanoid robots can display very complex behaviors, simpler systems such as 
microorganisms and microbots are the main target of a physics-based description and modelling of the interactions and emergent 
collective behavior in systems with a few or many intelligent active particles. However, humans, and  
larger animals in specific, controlled situations -- such as pedestrian streams meeting
at an intersection \cite{iyer2024_3way}, or bird flocks escaping from predators \cite{cavagna2014_annurev, cavagna_physics_2018} -- are also within the realm of such modelling approaches. 

As the field of intelligent active matter is broad and covers a wide range of systems and topics, we will focus here on the theoretical modelling of groups of motile 
intelligent particles, which can sense their environment and modify their motion by self-steering.

\section{Navigation}
A first challenge is the navigation of a smart active particle between two locations in a complex landscape with minimal travel time, energy dissipation, or fuel consumption \cite{colabrese2017_flownavigation, schneiderf2019_steering, liebchen2019_navigation}. A typical example of smart behaviour here is chemotaxis, which facilitates the navigation by bacteria toward nutrients, sperm cells toward the egg, and immune cells towards pathogens. Rule-based approaches include chemotaxis modeled as drift along chemical gradients \cite{liebchen2018_chemotaxis}, threshold switching in quorum sensing and gene regulation, and cellular-automata models based on local interaction rules. However, recent years  have also seen an increasing use of various machine learning techniques and genetic algorithms, which have shown to be very effective
in finding optimal solutions, even for the navigation of active particles in unknown environments with random obstacles of diverse shapes, sizes, and configurations \cite{yang2020_navigation,muinos2021_RL,gerhard2021_hunting}. Beyond navigation, swarms of individually controllable active particles can be used to capture and transport cargo via feedback control \cite{xu2021_mazes,stengele2023_cargo,heuthe2024_counterfactual,swarm-bot2006_short}.

\section{Pursuit and Evasion Dynamics}
\label{sec:noisy pursuit}

In the course of animal evolution, the development of sensory
organs has enhanced the precision
of sensing, thereby facilitating sophisticated predation schemes. Consequent competition among species implies high evolutionary selection pressure, the so-called arms-race
hypothesis. 

While the pursuer has to steer its motion toward the target
location, the evader, which is typically slower but more agile, has to compensate the speed disadvantage by exploiting information about the instantaneous pursuer location and velocity to develop promising escape strategies.
Clearly, noise and randomness play an important role for both
pursuer and evader \cite{meyer2024_nonmarkovian}. 

The steered motion of the pursuer with orientation vector ${\bf e}_p$ can be described by \cite{goh2022_noisypursuit}
\begin{equation}
\dot{\bf e}_p = {\bf e}_p \times \left[ \Omega_p ({\bf u} \times {\bf e}_p) + {\bm \Lambda} \right] ,
\label{eq:Langevin_pursuer}
\end{equation}
where ${\bf u} = ({\bf r} - {\bf r}_p)/|{\bf r} - {\bf r}_p|$
is the target direction in the reference frame of the pursuer, $\Omega_p$ is the pursuer maneuverability, and ${\bm \Lambda}$ is a Gaussian and Markovian stochastic process with zero mean and variance determined by the rotational diffusion coefficient $D_R$. The target can be stationary, as in sperm searching for the egg \cite{jikeli2015_helical_short}, move on a straight trajectory, or perform a run-and-tumble motion. Here and below, all maneuverabilites $\Omega$ are measured in units of $D_R$ and are therefore dimensionless.

In the absence of noise, the pursuer performs a deterministic motion around a stationary target, which consists of an elliptical trajectory with a perihelion rotation, determined by the initial conditions.  With noise,
pursuer propulsion speed $v_p$ is characterized by the 
P{\'e}cet number $Pe=v_p/(\sigma D_R)$, where $\sigma$ is the agent diameter and $D_R$ the rotational diffusion coefficient. In this case, (non-interacting) pursuers form a buzzing cloud around the target,
where the average distance $\langle r \rangle$  from the target scales as $(Pe/\Omega_p) F(\sqrt{D_R})$, with $F(x) = 1/x^2$ for $x<1$. Thus, noise is essential for the pursuer to reach the target \cite{goh2022_noisypursuit}.

\begin{figure}
\onefigure[width=0.85\columnwidth]{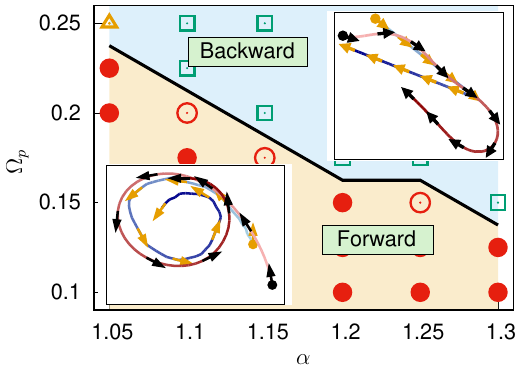}
\caption{Optimal tumbling strategy regimes, maximizing the capture time, for forward (large alert distance) ($\circ$, $\bullet$) or backward (small alert distance) ($\square$, $\triangle$) tumbling, as a function of pursuer maneuverability $\Omega_p$ and speed excess $\alpha=v_p/v_t$ (with target speed $v_t$).
Insets show representative trajectories for pursuer (red) and target (blue), with the color becoming progressively darker with time. Arrows indicate instantaneous pursuer (black) and target (brown) orientations.
Adapted from \cite{goh2026_runtumble}. CC BY 4.0. 
}
\label{fig:PD_RTP_escape}
\end{figure}

Noise and randomness is even more important for the evader,
as it makes its motion unpredictable for the pursuer 
\cite{bern22_hare, bernardi2024_random, goh2026_runtumble}.
Here, it is important to note that random motion of the evader is actually detrimental, as a diffusive motion effectively localizes the target \cite{bern22_hare}. For an evader, which recognizes on the position, the speed, and the direction of motion of the pursuer, the obvious choice is a run-and-tumble motion, where the tumbling rate depends on the instantaneous distance between them. Furthermore, the preferential tumbling direction can be adjusted to the 
direction of motion of the pursuer. Results then depend on the ``awareness range" of the target, namely, the average distance at which the need for tumbling becomes urgent, and the preferred angular tumbling range. As shown in Fig.~\ref{fig:PD_RTP_escape}, two beneficial escape scenarios can be distinguished \cite{goh2026_runtumble}. For predators that are only slightly faster and have a low maneuverability, forward tumbling is advantageous. This is surprising, as continuing on a straight motion seems optimal, however, the predator could never be outrun in this way. Thus, by a self-organized tumbling away from the predator (resulting from increasing tumbling frequency with decreasing distance), both agents move on quasi-circular trajectories, with the predator on the longer, outside track. For significantly faster predators with higher maneuverability, backward tumbling -- although risky -- is advantageous, as it gives a head-start for further escape by maximizing relative speed.

\section{Swarming}
\label{sec:swarming}

Three classical models have been suggested to describe the 
behavior of animal groups and herds: the ``boid" model \cite{reynolds1987_boids}, the Vicsek model \cite{vicsek1995novel, chate2008_vicsek} and the behavioral zonal model \cite{couzin2005_zonal}. Here, the first and last model 
incorporate short-range avoidance, medium range motion alignment,
and long-range group following, while the Vicsek model focuses
on alignment only.

\begin{figure}
\onefigure[width=0.88\columnwidth]{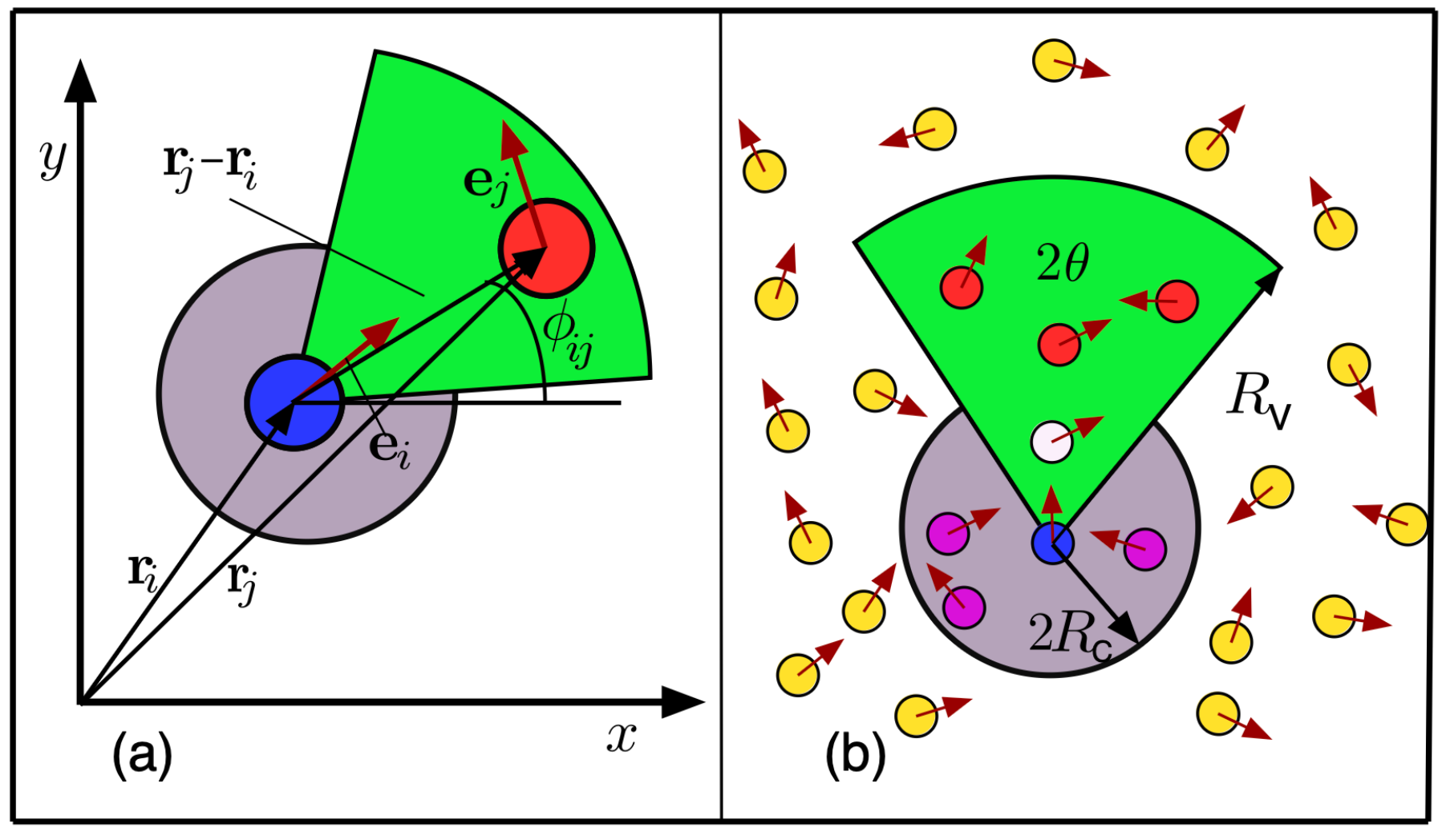}
\caption{(a) Schematic representation of vision cone and alignment neighborhood of particle $i$ (blue) with orientation ${\bf e}_i$, distance vector ${\bf r}_{ij}={\bf r}_j - {\bf r}_i$ to other particles. 
(b) Polar orientation field (grey) with cutoff $R_c$, and vision cone (green) with vision angle $\theta$ and vision range $R_v$. 
 Reproduced from \cite{negi2024_vision_align}. CC BY 4.0.}
\label{fig:iABPs_schematic}
\end{figure}

\subsection{Models: Vision- and alignment-induced self-steering}

The collective behavior resulting from perception-dependent motility has
been studied in more detail recently \cite{barberis2016_cognitive, negi2022_vision, goh2022_noisypursuit,  negi2024_vision_align, bastien2020_vision}. Directional sensing is taken into account in terms of visual perception with a vision cone of opening angle $2\theta$ and vision range $R_v$.
The main idea is that particles, denoted as intelligent active Brownian particles (iABP), move with
constant speed $v_0$, but redirect their motion according to the number and position of other particles in their vision cone. This redirection is described by a 
vision-induced torque, such that the orientation vector ${\bf e}_i$ of particle $i$ varies as
\begin{equation}
\dot{\bf e}_i = {\bf e}_i \times \left\{ \left[ \textbf{M}_{v,i} + \textbf{M}_{a,i}^{\alpha}\right] \times {\bf e}_i + {\bm \Lambda}_i \right\},
\label{eq:Langevin}
\end{equation}
where $\textbf{M}_{v,i}$ and $\textbf{M}_{a,i}$ are
the vision- and alignment-related adaptive generalized forces, respectively, and ${\bm \Lambda}_i^\alpha$ represents Gaussian and Markovian stochastic processes, as in Eq.~\eqref{eq:Langevin_pursuer}. Furthermore,
\begin{align}
\textbf{M}_{v,i} =& \frac{\Omega_v}{N_{v,i}}
\sum_{j \in VC} \exp(-r_{ij}/R_0) \ {\bf u}_{ij} \ , \ 
\label{eq:torque_visual} \\
\textbf{M}_{a,i} =& \frac{\Omega_a}{N_{a,i}} \sum_{j \in AR} {\bf e}_j,
\label{eq:torque_aligment}
\end{align}
where ${\bf r}_{ij} = {\bf r}_j - {\bf r}_i$ is the distance vector between particles `i' and `j', $r_{ij} = |{\bf r}_{ij}|$, 
and ${\bf u}_{ij} = {\bf r}_{ij}/r_{ij}$. The numbers $N_{v,i}$ and $N_{a,i}$ are the effective numbers of particles in the vision cone $VC$ and in the alignment range $AR$ of radius $R_c$, respectively. An excluded-volume interaction between particles can be taken into account by a short-range repulsion with particle diameter $\sigma$. It is important to notice that $\Omega_v>0$ favors steering toward regions of higher target density, and thereby cohesion and clustering, while $\Omega_v<0$ favors steering away from the target, and thereby neighbor avoidance.

\begin{figure}
\onefigure[width=0.95\columnwidth]{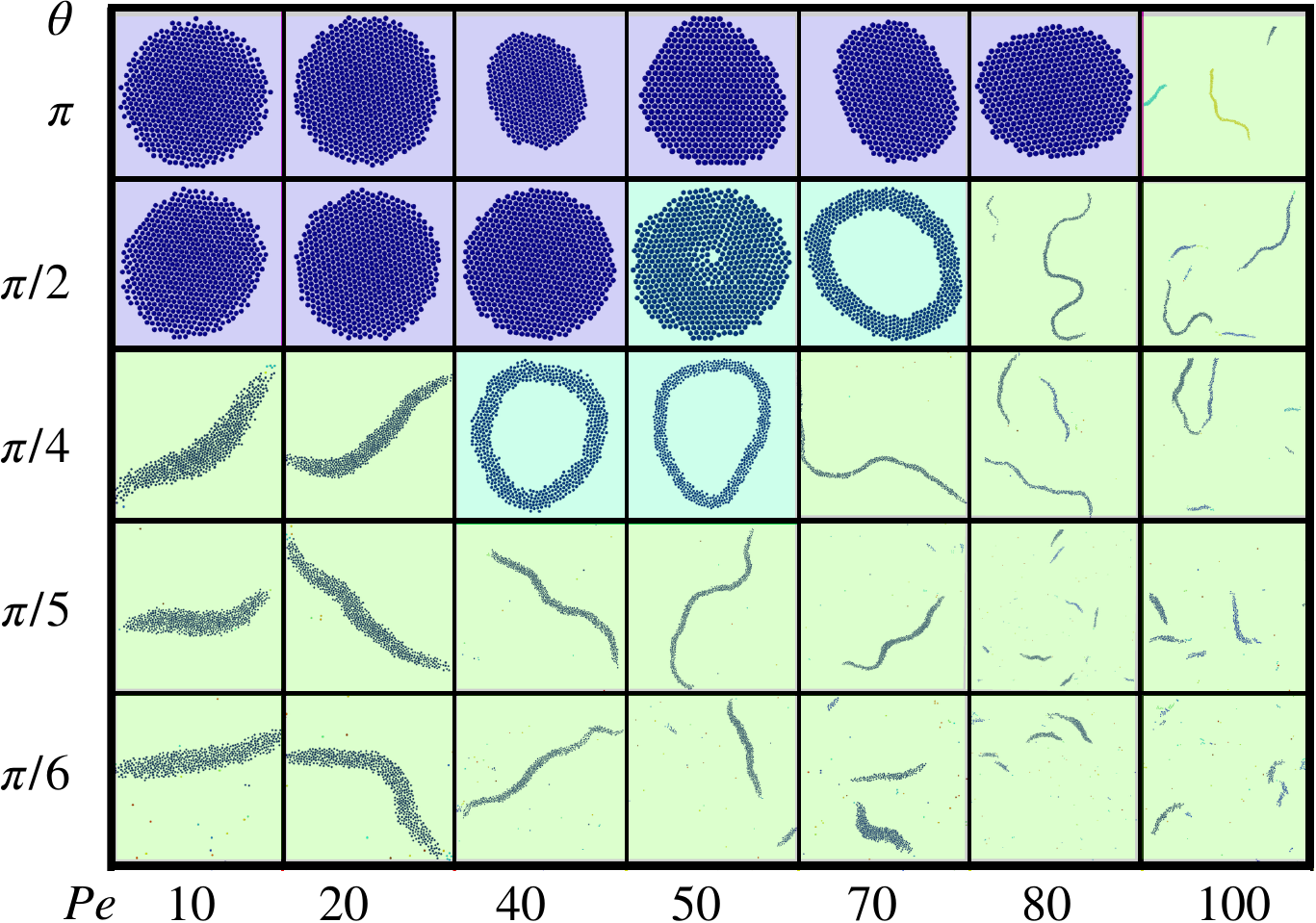}
\caption{Snapshots of emerging structures for different P\'eclet numbers $Pe$, vision angles $\theta$, with alignment-vision ratio $\Omega_a/\Omega_v = 4$ and packing fraction $\Phi= 0.00785$. The snapshots are not to scale for better visualization. 
Reproduced from \cite{gompper2025_roadmap_short}. CC BY 4.0.
}
\label{fig:vision_alignment}
\end{figure}

\subsection{Non-reciprocity}
Interactions in this minimal cognitive model are inherently nonreciprocal, because: (i) directed sensing in the vision cone, where one agent might see
another, but not vice versa; (ii) the asymmetry in the steering torques already at the
two-particle level, even for panoramic view (vision angle $\theta= \pi)$; and (iii) for binary systems, in particular predator-prey systems, the different reaction of A-agents to B-agents compared of B- to A-agents, typically reflected in the different maneuverabilities $\Omega^{A,B} \ne \Omega^{B,A}$.

\subsection{Vision-induced activity}
Another type of perception-dependent collective behavior has been
suggested for colloidal microrobotic systems \cite{laverge2019_perception}. Here, particles become activated 
(propulsive) when the effective number of other particles in the vision cone exceeds a threshold. This implies that particles at the
perimeter of a cluster are active/passive if their orientation is toward the inside/outside of the cluster, thereby enhancing cluster formation and stability. In case of misalignment of self-propulsion direction and perception cone axis, this model predicts a persistent swirling motion
of clusters with well-defined chirality \cite{saavedra2024_misaligned}.

\subsection{Swarm behavior}
Swarm formation occurs due to the local motion alignment and 
cohesive group dynamics of many identical individuals \cite{barberis2016_cognitive, bastien2020_vision, negi2022_vision, negi2024_vision_align, shea2025_cohesive, liu2025_vision3D}.
The iABP model defined by Eqs.~(\ref{eq:Langevin}), (\ref{eq:torque_visual}) and (\ref{eq:torque_aligment})
predicts that swarm shape and dynamics depends strongly on
particle speed $v_0$, characterized by the P{\'e}clet number $Pe=v_0/(\sigma D_R)$, the competition between alignment and cohesion, determined by the (dimensionless) maneuverabilities $\Omega_v$ and $\Omega_a$, and the vision angle $\theta$.  

A section of the phase diagram, for fixed $\Omega_a/\Omega_v=4$ is shown in Fig.~\ref{fig:vision_alignment}. For large vision angle $\theta$ and small P{\'e}clet numbers, a significant number of neighbors occupy the vision cone and steering is efficient since $\Omega_v \ge Pe$, both of which favor the formation of large clusters. In contrast, for small vision angles and large 
P{\'e}clet numbers, particles focus on the single particle in front, and their propulsion dominates over orientational noise, which favor the formation of thin elongated swarms, possibly even single-file motion \cite{barberis2016_cognitive, negi2022_vision}. Alignment steering also contributes to swarm elongation \cite{negi2024_vision_align}.

\subsection{Information propagation and behavioral inertia}

\begin{figure}
\onefigure[width=\columnwidth]{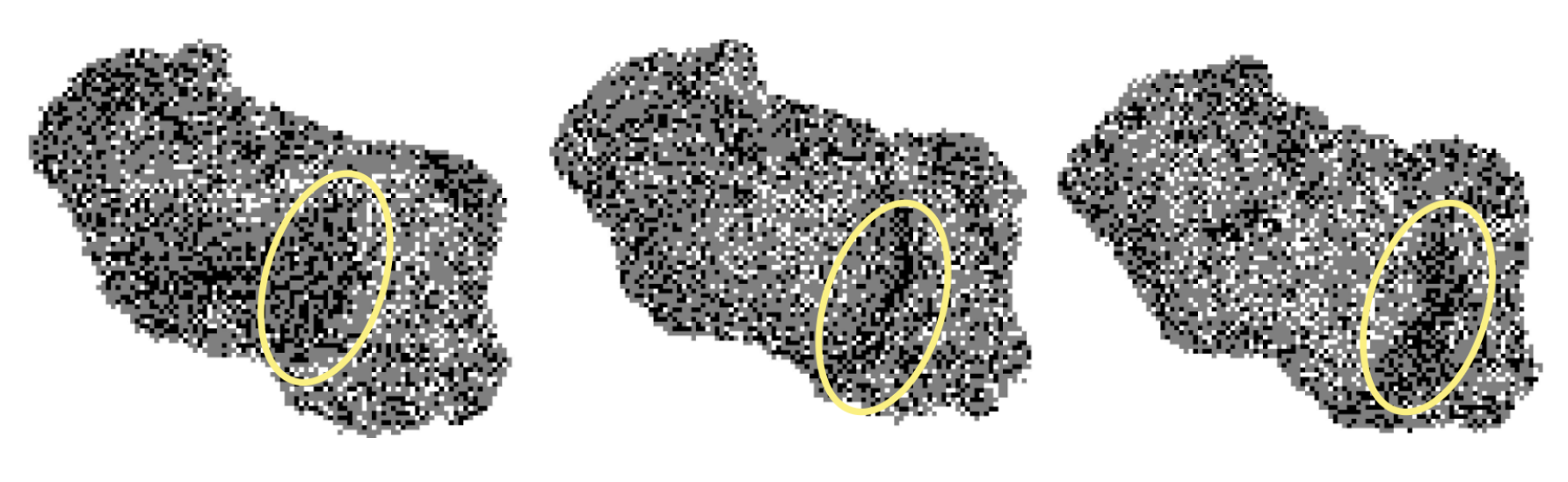}
\caption{Flock snapshots with progressing time (left to right), 
demonstrating the propagation of large-scale density fluctuations (emphasized by yellow ellipse). Reproduced from \cite{iyer2026_edge}. CC BY 4.0.}
\label{fig:ISM_waves}
\end{figure}

For some bird flocks, experimental observations provide strong evidence that information propagation is not diffusive (as in the Vicsek model), but propagates with finite speed
\cite{cavagna2015_turning_short, cavagna_physics_2018}. In order to explain this behavior, Cavagna and Guardina proposed the inertial spin model (ISM), an extension of the Vicsek model,
where each particle carries a spin $s_i$ with mass
\cite{cavagna2015_turning_short, cavagna_physics_2018}.
This model can be generalized to also include avoidance steering at short distances, and ``following" steering for group cohesion \cite{iyer2026_edge}. The equations of motion then become
\begin{equation}
\label{eq:ISM}
  \begin{aligned}
    & \dot{\textbf{e}}_i =  \textbf{s}_i \times  \textbf{e}_i, \\
    & \chi \dot{\textbf{s}}_i = \textbf{e}_i \times \left [ -\eta \textbf{s}_i + \boldsymbol{\zeta}_i + \textbf{M}_i \right ],
  \end{aligned}
\end{equation}
where the adaptive generalized forces for visual following and avoidance, Eq.~\eqref{eq:torque_visual}, and for alignment, Eq.~\eqref{eq:torque_aligment}, are the same as in the iABP model. 
In Eq.~(\ref{eq:ISM}), $\chi$ is a generalized
rotational (behavioral) inertia, $\eta$ a damping coefficient, and the noise term  $\zeta$ a Gaussian random process with zero mean and variance controlled by a temperature $T$.
When the steering for group cohesion is only active at the swarm boundary, then this model reproduces the linear information propagation of the original ISM. Furthermore, it displays several interesting behaviors (in two spatial dimensions) \cite{iyer2026_edge}: 
(i) An order-disorder transition between
disordered and polar flocks, which is {\em not} in the Vicsek universality class. 
(ii) A flock shape, which is compact and roughly circular for 
high cohesion maneuverability $\Omega_v$, but branched and irregular for small $\Omega_v$; an analysis of shape fluctuations suggests the emergence of an effective line tension $\lambda$, with $\lambda\sim \Omega_v$. 
(iii) The formation of density waves inside the flock, see Fig.~\ref{fig:ISM_waves}, which are reminiscent of the murmurations of bird flocks in the sky. Beyond inertia, rapid information propagation may be facilitated by other mechanisms such as a natural time-delays in the response \cite{holubec2021_timedelay} or by immediate and explicit visual feedback of information such as a change in body orientation \cite{zheng2024_scalefree}.

\section{Collective Predation and Escape}

The swarming model, defined by Eqs.~(\ref{eq:Langevin}) and (\ref{eq:torque_visual}) can be easily generalized to a binary system 
of $A$ and $B$ particles \cite{negi2025_binary}. For the vision-induced generalized steering force, this implies
the generalization
\begin{equation}
\textbf{M}_{v,i}^{\alpha} = \sum_\beta \frac{\Omega_v^{\alpha\beta}}{N_{v,i}^{\alpha\beta}} 
\sum_{j \in VC} \exp(-r_{ij}/R_0) \ {\bf u}_{ij},
\label{eq:torque_visual_binary}
\end{equation}
with $\{\alpha, \beta\} \in \{A,B\}$.
The main difference to single-component systems is that steering now involves four different maneuverabilities, where
in particular $\Omega_{AB} \ne \Omega_{BA}$, even with different signs. 

\begin{figure}
\onefigure[width=\columnwidth]{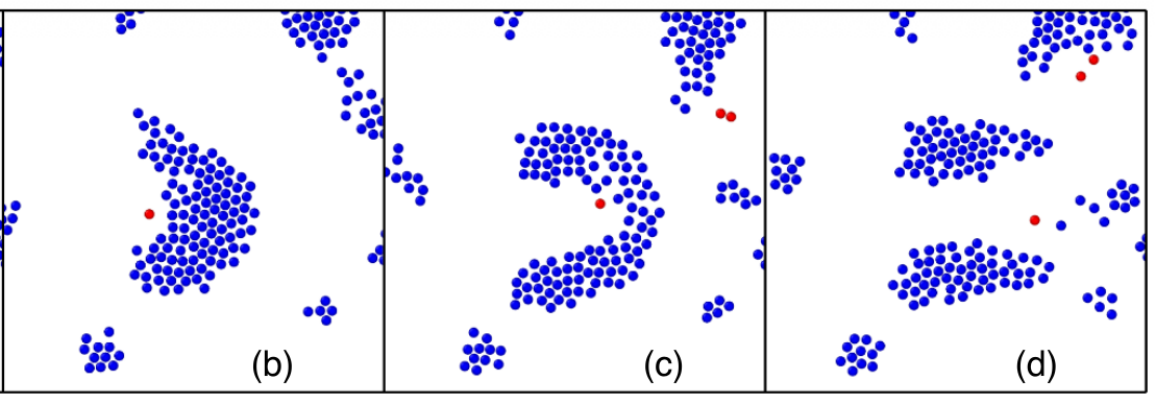}
\caption{Snapshots of a predator-prey pursuit, progressing in time from left to right, and illustrating a typical pursuit of a predator (red) chasing prey particles (blue). Reproduced from \cite{negi2025_binary}. CC BY 4.0.}
\label{fig:binary_predator_prey}
\end{figure}

This is the case for the typical predator-prey scenario, where $A$ follows $B$, but $B$ avoids $A$~\cite{angelani2012_strategies}. A typical pursuit is shown in Fig.~\ref{fig:binary_predator_prey}.
The simulations demonstrate that the vision angle $\theta_A$ of the predator plays an important role \cite{negi2025_binary}. For small $\theta_A$, the narrow visual field implies a reduced prey detection and subsequently low prey densities in front of the predator. As $\theta_A$ increases to about $\pi/4$, which provides focused vision (‘eagle’s eye’), the prey density in front of the predator reaches its maximum. The broader field
of view for $\theta_A \ge \pi/2$ counteracts an effective chasing of the prey, because the simultaneous visibility of
prey in many different directions implies a less goal-oriented and more erratic motion of the predator.

\section{Pedestrian Crowds}

Pedestrians are a very special type of cognitive active agents, as they can not only avoid, follow, or align with neighboring particles, but also follow a prescribed goal direction, for example head to the other side of an intersection. In counter-flow motion, collision avoidance and goal orientation lead to ``laning" \cite{zhang2012_laning}. 

An interesting reference situation is pedestrians moving with roughly constant speed at fixed average density. In order to keep their preferred social distance, or to avoid potential infections (COVID-19!), they have to constantly change their
direction of motion 
\cite{echeverria_2021_social_distancing_short, negi2024_controlling}. Simulations of the model \eqref{eq:Langevin} with
avoidance steering ($\Omega_{v}<0$) show, for example, that probability distributions for the average nearest-neighbor distance $d_1$ for constant ratio $Pe^{3/2}/\Omega_v$ collapse  onto a single master curve, which indicates a tight coupling of individual activity and maneuverability.

\begin{figure}
\onefigure[width=0.99\columnwidth]{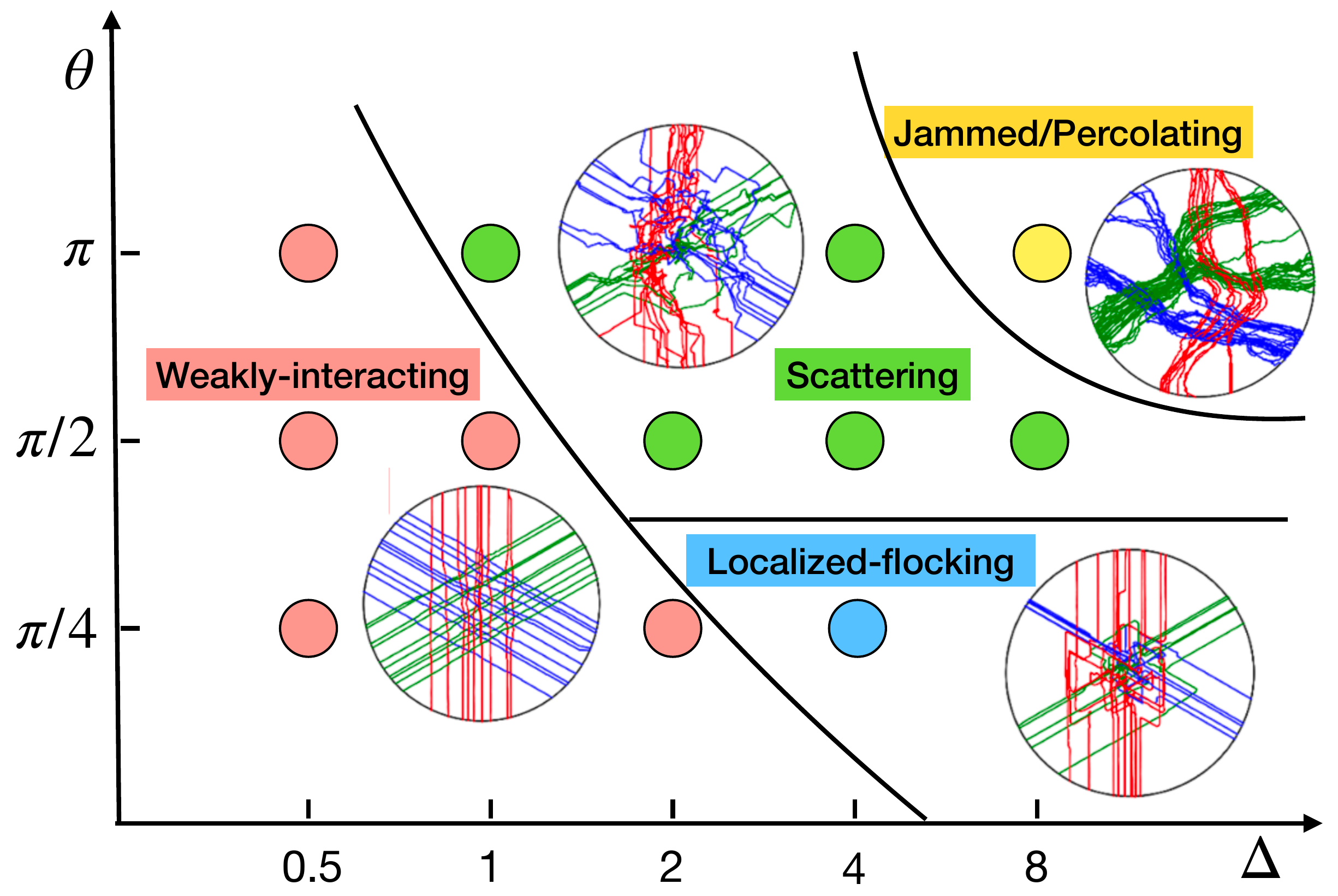}
\caption{State diagram of pedestrian
movement states in a three-way intersection, as a function of the relative maneuverability $\Delta= \Omega_v/K$ and vision angle $\theta$, where $K$ is the alignment  of each agent toward their goal. For small $\Delta$, agents pass
through the interaction zone nearly unhindered (weakly-interacting), for intermediate $\Delta$ and vision angles $\theta \ge \pi/2$, agents attempt to avoid each other while crossing. For high $\Delta$ and large $\theta = \pi$, a jammed, percolating phase develops. Intermediate $\Delta$ and small $\theta < \pi/2$ yields a ‘localized flocking’ regime, where agents
align with oncoming individuals, forming a local co-moving pedestrian cluster. Adapted from \cite{iyer2024_3way}. CC BY 4.0.
}
\label{fig:PD_three_way_crossing}
\end{figure}

A paradigmatic scenario for the directed motion of agents in semi-dense crowds
(where neighbor distances are sufficiently large that excluded volume interactions are not important) is the interaction of several streams of active agents with goal orientation at a three-way intersection \cite{iyer2024_3way}. The resulting phase diagram is shown in Fig.~\ref{fig:PD_three_way_crossing}, and displays several types of collective behavior,
including localized flocking, jamming and percolation, and self-organized rotational flows. Despite of this
variety of behaviors, the fundamental flow diagram shows a universal curve for different
vision angles \cite{iyer2024_3way}.

Another interesting case is the directional motion of a self-steering active intruder in a crowd of cognitive active agents
\cite{nicolas2019_pedestriancrowds, kushwaha2026_intruder}. Simulations show that the intruder’s attempt to increase directional
speed by steering around agents fails -- in fact, this even reduces the directional speed. In contrast, the intruder has to be perceived by the agents so that they can move out of the way in time.

\section{Hydrodynamics}

An important distinction in active matter is between ``dry" and ``wet" systems. The reason is that the behavior depends on the type of hydrodynamic propulsion, which is related to hydrodynamic interactions and generated ``active fluid stress", which falls into the three main categories of pusher, neutral swimmer, and puller. These microswimmers can be modelled on a generic, coarse-grained level as ``squirmers", with an imposed fluid flow field on their surfaces. Steering is facilitated by non-axi-symmetric surface flows \cite{pak2014_steering, goh2023_hydropursuit}. In a pursuit situation with pushers, faster pursuers cannot easily catch up with slower evaders, as there is a hydrodynamic repulsion
between them, which acts to equalize the velocities.
For pursuers aiming to align their propulsion direction with the evader, the formation of stable doublets requires speed adaptation, because the pusher flow field of the pursuer exerts a torque on the evader, which turns it away. Thus, the pursuer has to take the longer ``outside line", and has to compensate by speeding up \cite{goh2023_hydropursuit}. Hydrodynamics can of course also play an important role in microswimmer navigation
\cite{daddi-moussa-ider2021_navigation}. 

For a suspension of many self-steering and aligning microswimmers -- the wet Vicsek model --,  the behavior strongly depends on the propulsion type.
Pushers have a homogeneous density distribution and 
display active turbulence, while pullers exhibit the intermittent formation of clusters, and vortex rings \cite{goh2025_hydrovicsek}.

\section{Future perspectives}

The current state of the study, understanding, and design of intelligent active systems promises an exciting
future, in which active matter and distributed intelligence converge not only
to understand biological systems, but also to develop artificial (micro)systems \cite{chung2011_robots, ju2025_Technology_short, janzen2026_robophysics, miskin2026_robots} capable
of sensing, learning, and collectively solving tasks. Understanding local sensing and decision-making is a crucial step toward distributed computing. Here,
groups of agents can be made to transport cargo, attain a goal, or ‘resolve’ traffic
jams without the need for central control. In this regard, we can take inspiration
from natural systems such as ants and bees, applying aspects of swarm intelligence so that groups of agents with minimal ``brains” can jointly execute complex tasks. \\
Here, challenging issues for future research include: \\
{\em Noise and minimal brain --} There are several sources of noise in intelligent active systems, such as thermal motion, active (motor) noise, errors in sensing the environment, noise in decision making due to incomplete information. Weak noise might help to attain
a goal, but at some level it becomes detrimental. 
At the microscale, despite a fuzzy and
noisy environment, organisms sense and steer using molecular signalling pathways — for instance, a neutrophil chasing a bacterium. This poses natural optimization and design challenges in terms of the control of microscale robots which can only be engineered with limited capabilities. Externally feedback-controlled systems can bypass some of these limitations~\cite{muinos2021_RL,heuthe2025_swarmalator}, as the sensing and processing of individual agents happen externally, mediated via dynamic image processing and control. However, a key challenge remains the developed of autonomous agents, akin to microswimmers, that can operate and deliver cargo without relaying on an external 'brain'. Here the ``intelligence" of each agent will be limited by the balance of resource availability and task performance.


{\em Species-specific modelling --}  The swarming and pursuit behavior of different animal species can be wildly different. Not only are fish schools different from bird flocks and bee swarms,
but also starlings behave differently from geese, etc. Thus, a single theoretical model should not be expected to capture the dynamics and self-organization across many species. Therefore, a hierarchy of models is required, from the generic models of agents 
with simple sensing and decision-making rules, to increasingly specific models for individual species. Species-specific studies have been performed, and models developed,
for example for locusts \cite{sayin25_locust_short}, ants \cite{gelblum2015_ants}, and sheep \cite{gomez2022_alternating}. Such studies can help to identify the main interactions
that guide an interesting specific behavior. For instance, the generalised ISM model to understand information propagation is suitable for describing information transfer in groups of flocking birds without a leader. However, for foraging groups of birds that are typically stationary, an approaching threat leads to a ``fear"-induced flight response that spreads rapidly in the group in a relay-like propagation and is not captured by the ISM. Likewise, different animal groups interact through metric \cite{evangelista2017three}, topological \cite{ballerini2008_topological_short}, or visual \cite{wirth_2023_PNASnexus} neighbourhoods, and therefore exhibit different responses to changes in density. 
Furthermore, perception is not always visual,
but can involve sound \cite{ziepke2025acoustic} and sonar \cite{astrup1999_ultrasound}, as well as chemotatic sensing \cite{raina2019_symbiosis, liebchen2018_chemotaxis}.
Thus, theoretical modelling requires an intersection between physics and biology to arrive at simple yet sufficiently complex models that can capture the underlying mechanisms. In particular, engineering of diverse models can help uncover several routes to intelligence and emergence inspired by nature.

{\em Nonreciprocal interactions--}
Non-reciprocity of interactions between cognitive active particles are the rule rather than the exception and motivates the study of a broad new class of nonequilibrium phenomena. Indeed, non-reciprocity 
has been identified as the source of wide range
of novel dynamical behaviors and self-organization
\cite{fruchart2021_nonreciprocity}, like asymmetric clustering in active mixtures induced by nonreciprocal alignment \cite{kreienkamp2024_nonreciprocal_alignment}, 
the emergence of collective chiral motion in aligning/anti-aligning binary systems \cite{chen2024_chirality},
and the spontaneous emergence of time-dependent states that break 
parity-time symmetry \cite{kreienkamp2025_exceptional}. 

{\em Response adaptation --} Across scales, from bacteria to fish to humans,
groups of agents not only interact with each other but also adapt their responses
to changing environments. A simple example is the behaviour of a pedestrian
in a relaxed setting, such as strolling, versus a stressed or panicked environment,
such as during an evacuation. A behaviour of adaptive task sharing has been studied in small groups of grazing sheep \cite{gomez2022_alternating}. Similarly, bacteria adjust their motion, metabolism,
and reproduction rate depending on resource availability, environment, confinement, and other factors. In such scenarios, a single model cannot capture the vast
array of behaviours observed. 

{\em Swarm intelligence --} The idea of swarm intelligence is that the swarm is more intelligent, and can perform more complex tasks, than the individual agent. The usual examples for swarm intelligence are ants and bees. In both cases, individuals communicate about the location and access to food sources, in case of ants by pheromone trails, in case of bees by complex ``dances". Also, ants have developed sophisticated strategies to transport large cargo collectively \cite{gelblum2015_ants, dreyer2025_puzzle}. 
It is certainly interesting to pursue similar approaches for the design of microbots.
It would also be interesting to find and utilize other types of
swarm intelligence, such as using slime
mould (\emph{Physarum}) for network optimzation \cite{tero2010_rules_short}, 
or more explicitly exploiting stigmergic interactions mediated by persistent environmental memory~\cite{theraulaz1999stigmergy,dias2023_environmental}. Lastly, intelligent swarm collectives must be capable of performing tasks based only on information from their local neighbourhood. Here, a key challenge involves developing coordination strategies that can work with little to no information about the global state, while also ensuring scalability to larger systems \cite{janzen2026_robophysics}.

{\em Machine learning and artificial intelligence --}
Artificial intelligence can steer complex, noisy systems toward desired target states, such as cargo delivery or collective shape formation, by handling nonlinear, high-dimensional dynamics that are difficult or impossible to optimize using analytical methods. 
In fact, optimal control theory and reinforcement learning have already employed in a variety of microscale systems
\cite{krishnan2026hamiltonian, durve2020_learning, yang2024_ML4Bots, cichos2020_ML4AM, heuthe2024_counterfactual, paul2025_embodiment} to achieve precise tasks. Such approaches will receive
rapidly increasing attention, in particular for systems with complex
perception and a spectrum of activities.

{\em Conclusions --} There is much to be learned and understood from nature. Evolution
has selected certain emergent behaviors that allow groups of simple agents
to outperform their individual cognitive capabilities. Often, these behaviours
rely on some level of sensing and response, which may differ in ``intelligence,”
but share the overarching goal of developing a strategy that may directly (like
avoiding threats) or indirectly (building complex nests) improve survival. 
The combination of theory, simulations, and experiments, will facilitate a better 
understanding of these diverse behaviors, but also the transfer of knowledge 
into engineered systems that can achieve remarkable collective feats.

\acknowledgments
Discussions with  
C.~ Bechinger, F.~Cichos, D.A.~Fedosov, R.S.~Negi, M.~Ripoll, 
A.~Schadschneider, S.P.~Singh, and R.G.~Winkler are gratefully acknowledged.

\bibliography{intelligence2.bib,gompper_newlabels.bib}
\bibliographystyle{eplbib.bst}

\end{document}